# A three-year comparative study of dominant misconceptions among first-year physics students at a South African university

Anna Chrysostomou[1,2], Alan S. Cornell[1], and Wade Naylor[3]

[1]Department of Physics, University of Johannesburg, PO Box 524, Auckland Park 2006, South Africa
[2]Institut de Physique des Deux Infinis de Lyon, Université de Lyon, UCBL, UMR 5822, CNRS/IN2P3, 4 rue Enrico Fermi, 69622 Villeurbanne Cedex, France
[3]National School of Education, Faculty of Education and Arts, Australian Catholic University, Brisbane, Australia

This article discusses a three-year study (2020 − 2022) of dominant misconceptions (DMs) for a large cohort of first-year physics course students at the University of Johannesburg (UJ), South Africa. Our study considered pre-test scores on the *Force Concept Inventory* using a graphical method, where we found statistical differences between the mean DM scores for the 2020 cohort, as compared to the 2021 and 2022 cohort; possibly due to the onset of COVID lockdowns. We also compared our data from South Africa with cohorts based in Spain and the Kingdom of Saudi Arabia, where the method of DMs was also applied. From this comparison, we found some differences in the preconception knowledge of the cohorts. Furthermore, we included an analysis of DMs through the 'gender lens' for the South African cohort, finding no statistically significant difference between the means for DM scores of students who identify as male or female. Finally, given the diverse language backgrounds and levels of matriculation preparation for university level physics courses, we have also shown how quickly responding to student misconceptions can be efficiently addressed using the method of DMs.

Wade Naylor E-mail: wade.naylor@acu.edu.au (author for correspondence)
Alan S. Cornell E-mail: acornell@uj.ac.za
Anna Chrysostomou E-mail: achrysostomou@uj.ac.za
The authors have followed the Contributor Roles Taxonomy:
Anna Chrysostomou - Conceptualisation, Methodology, Investigation, Data Curation, Writing - review & editing;
Alan Cornell - Conceptualisation, Methodology, Validation, Writing - review & editing, Funding acquisition;
Wade Naylor - Conceptualisation, Methodology, Formal analysis, Investigation, Visualisation, Project administration, Writing - original draft, Funding acquisition.

## Introduction

One cannot understate the importance of assessment in all its forms and guises; to quote Black and Wiliam (1998), "We use the general term 'assessment' to refer to all those activities undertaken by teachers − and by their students in assessing themselves − that provide information to be used as feedback to modify teaching and learning activities". The National Council of Teachers of Mathematics [NCTM], 1995, for example, states that assessment is "the process of gathering evidence about a student's knowledge of, ability to use, and disposition toward mathematics and of making inferences from that evidence for a variety of purposes" (NCTM, 1995). Assessment in 'physics' is of course of necessary importance as outlined in Main (2022) and references therein. One such type of assessment inventory used in Physics Education Research (PER) is the *Force Concept Inventory* (FCI) (Halloun & Hestenes, 1985; Hestenes, 1998; Hestenes et al., 1992). The FCI has many uses and applications including in 'assessment for learning' (Gioka, 2006)) and the study of Learning Progressions (Fulmer et al., 2014), which are of extreme importance in guiding curricula in various countries (see, for example, the reports from the Australian Curriculum and Assessment Reporting Authority (ACARA, 2017) and National Research Council (NRC, 2012)).

The FCI has been used extensively as a 'summative assessment' tool, such that instructors deploy the test at the end of a semester in order to evaluate student understanding of the concepts taught in an introductory classical mechanics module. However, it can also be administered as a 'formative assessment', where students are asked to complete the FCI without prior knowledge of the testing, in order to establish a 'snapshot' of each student's conceptual understanding at the moment of testing. This can be a useful means of identifying misconceptions and is in line with the FCI as being used for more than just summative assessment (Stoen et al., 2020).

One such approach to get 'formative feedback' using the pre-test stage of the FCI is that of 'dominant misconceptions' (DMs). Note that there are only a few articles that explicitly use DMs, where the phrase 'dominant misconceptions' appears to have been coined by Martín-Blas et al. (2010). In particular, Martín-Blas et al. (2010) and Bani-Salameh (2016a) have argued that conceptual flaws in understanding should be diagnosed as early as possible, by analysing results at the 'pre-test' stage. For those students who choose a particular 'incorrect answer', this may unwittingly reveal that they suffer from one of the commonly held misconceptions outlined in Bani-Salameh (2016a, 2016b) and Martín-Blas et al. (2010), which are referred to as 'DMs'.

It is worth reemphasising that the reason for using the FCI as a 'formative assessment' tool (Black & Wiliam, 1998) is to provide 'immediate feedback' that can also be used as 'assessment for learning' (Gioka, 2006; Main, 2022) in the context of conceptual physics understanding. The FCI has also recently been used in physics to gain insights into the emotions influencing student performance during assessment (Lee et al., 2021) which leads directly to 'adaptive teaching'. This is where the teacher is guided by 'assessment for learning' (Gioka, 2006) to make decisions on how to proceed with teaching. In regards to alleviating anxiety, this is closely related to 'self-determination theory' (SDT) (Ryan & Deci, 2000), where learners are made aware of their motivations via this immediate feedback strategy (Lee et al., 2021). This in turn links to the concept of 'self-efficacy' (Bandura, 1977; Zimmerman, 2000) where recent research by Singh and colleagues (Cwik & Singh, 2022; Malespina & Singh, 2022) has investigated how students perceive their own academic performance, and the relationship between test-anxiety and self-efficacy in the context of gender.[1]

In this regard, women demonstrate low self-efficacy scores when compared against their male counterparts, even in cases where the former outperform the latter (Marshman et al., 2018). Given the well-documented shortage of women in physics (Eddy & Brownell, 2016) and the persistence of entrenched gender stereotypes, this can be interpreted as a 'quantifiable measure' of 'imposter syndrome' (Marchand & Taasoobshirazi, 2013). With this in mind, we consider the gender context to be of paramount importance in research investigating equity and gender differences in physics, in particular in relation to the FCI test (Alinea & Naylor, 2017; Bates et al., 2013; Bouzid et al., 2022; Coletta, 2013; Docktor & Heller, 2008; Glasser & John P. Smith, 2008; Hewagallage et al., 2022; Madsen et al., 2013; Shapiro & Williams, 2012). In South Africa, where the population reflects a wealth of cultural and linguistic backgrounds (as we shall discuss later), this issue of gender diversity may be even more prevalent (Francis et al., 2019; Sá et al., 2020).

**Background**

We now to return to the issue of why PER is carried out in the first place. It has long been known (Fagen et al., 2002; Mazur, 1997) that although students may be able to perform calculations in physics correctly, this does not necessitate their 'conceptual understanding' of the subject. There are various reasons for these barriers in physics comprehension, be they related to 'misconceptions' and/or other issues relating to gender, demographics or self efficacy, e.g., see (Cwik & Singh, 2022; Marshman et al., 2018). One way to inform teaching practices in physics and hence highlight the mentioned barriers is to employ the FCI. However, rather than utilise the FCI in its 'summative' form (comparing 'pre-' and 'post-tests' to calculate a normalised gain), one can use information in the 'pre-test' answer choices to help inform teaching practices.

To help explain how we will use the 'pre-test', we will first discuss the normalised gain $G$ that is formally defined as (Hake, 1998):

$$G = \frac{\langle \%S_f \rangle - \langle \%S_i \rangle}{100 - \langle \%S_i \rangle}, \qquad (1)$$

and is the usual way of using the FCI, where $\%S_f$ and $\%S_i$ are the final and initial FCI test scores, respectively. There are of course well discussed issues with this definition of normalised gain. An obvious challenge is that we need to wait for the post-test results to analyse the data. Another issue is that Eq. (1) only looks at differences and does not distinguish in a gain from, say, a student who scored a 5 → 10 vs. a score of 25 → 30 from *pre* → *post*-test. This issue stimulated work using 'Item Response Theory' (IRT) (Morris et al., 2006; Wang & Bao, 2010) and the semi-related 'Rasch model' (Nitta & Aiba, 2019; Planinic et al., 2010), where every single item (question) is analysed in terms of a 'question difficulty parameter' and a 'student ability parameter'.[2]

For example, Nitta and Aiba (2019) use the Rasch model to define a mathematically consistent definition of gain (known as 'Rasch gain'). This has the pleasing aspect that the gain, as defined from the class averages, is equal to the average of the individual student gains – something the normalised gain $G$ does not satisfy. These methods have even made it possible to develop the idea of 'computerised adaptive testing' in the FCI (Yasuda et al., 2021). As the authors discuss, the usual 30-question / 30-minute FCI could be completed with just a fraction of those 30 questions, and the time taken. Other interesting work by Nissen et al. (2018) also discusses how the normalised gain, Eq. (1), compares with Cohen's *d*, which is commonly used in the social sciences for defining effect sizes.

---

[1]To reflect, the participants in this study were reported with gender as male or female, and we limit our discussion to these two genders.

[2]The Rasch model is identical to the one-parameter IRT model, but the philosophical justification for each of the models is different: the Rasch model is top-down, where those parameters are based on assumptions. The IRT, on the other hand, is a bottom-up approach where one chooses the best parameterisation of the data (Nitta & Aiba, 2019). Another early method (Bao & Redish, 2001) in the context of the FCI used 'concentration analysis' to look at the grouping of incorrect responses as well as correct responses.

Although IRT and Rasch models are very informative, they do require large samples and resource-intensive statistical tools to apply them, particularly at the start of a PER program. Instead of returning to the normalised gain (see Eq. (1)), which as we have discussed is not always ideal, an alternative and less data-intensive approach would be preferable. A conceptual understanding tool for first-year physics students, that can be used straight after the 'pre-test' would be useful at Higher Education Institutes (HEIs) around the world.

In recent work, Carleschi et al. (2022) used the 'misconception' lens[3] to look at how undergraduate students in a classical mechanics course performed on the FCI assessment (Halloun & Hestenes, 1985; Hestenes, 1998; Hestenes et al., 1992). Interestingly, they found that although the 2020 cohort was subjected to COVID-19 lockdown(s), the gains for that cohort were 'standard' ($G \sim 0.25$, see Eq. (1)), despite the various challenges that the students faced.[4]

In a similar vein, we have extended the FCI test for three years, where we have the interesting opportunity to compare the cohorts over this time-period to determine if the lockdown period has had any effect on the conceptual understanding − at the 'pre-test' phase. As we have already argued, the use of the 'pre-test' is important because it allows a lecturer / course coordinator / School or Department to collect a 'snapshot' of the 'pre-conceptions' of each cohort at the start of the course − allowing for those teaching to make 'informed decisions' on how best to deliver the content (Madsen et al., 2017). We will also compare DMs at UJ with those results from other countries. Also, given the large size of the three-year data ($N \sim 800$), we are also able to compare DMs for a large number of students who identify as male or female in the course, an issue which we address in this article.

With the above goals in mind, we will introduce an adaptation of the idea of 'persistent' or 'dominant' misconceptions generally held by incoming undergraduate students[5] as we will explain in the next section.

**Dominant misconceptions and the literature**

To help explain what we mean by a DM, we consider a multiple-choice question such as the question in Figure 1 with 5 possible answer choices: *A, B, C, D, E*. The following equation can then be used to calculate a DM for a given question:

$$DM = \%(\text{largest incorrect answer}) . \qquad (2)$$

For example, let us say the correct answer to question 12 of Figure 1 is *B*. Suppose 45% of a class chose B but 51% of the class chose the incorrect answer *A*. The largest incorrect response corresponds to the (answer motivated by a) 'dominant misconception': $DM = 51\%$. The particular choice of DM may give useful insights into the conceptual understanding of that particular cohort, where usually a DM is defined as a % incorrect choice above a threshold - such as 50% or 60%, see later. In our example here, we can infer from Path *A* that the (flawed) logic motivating this choice is that equal contributions from a force acting in the direction of motion and the downward force of gravity govern the trajectory of a projectile (see Tables A1 and A2).

Our preliminary investigations conducted with first-year physics students at the University of Johannesburg (UJ) in 2020 (Carleschi et al., 2022) appeared to support the idea of DMs. For example, one dominant incorrect response was that 'a force in the direction of motion' was partially responsible for the action described in the corresponding FCI question (we refer the interested reader to Bayraktar (2009, p. 274) for a more detailed discussion of the common force and motion misconceptions). The belief that 'motion requires an active force' was flagged as a misconception, as identified by Bani-Salameh (2016a, 2016b) and Martín-Blas et al. (2010). A review of the 'taxonomy' of the coding and concepts on the FCI can be found in Appendix A (supplemental), where Tables A1 and A2 can be found, and are based on the work of Bani-Salameh (2016a), Bayraktar (2009), and Hestenes et al. (1992).

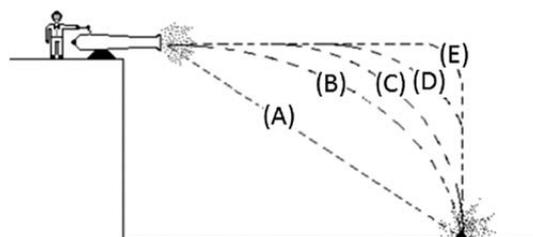

**Figure 1**

*Diagram for question 12 on the FCI, which provides the answer choices for the trajectory of a ball fired from the top of a cliff: A, B, C, D, E.*

---

[3]We will adhere to the phrase 'misconception'; however, it may be worth considering the use of other terms such as 'preconception' or 'anchor' (Clement et al., 1989), or 'alternate conceptions' (Maries & Singh, 2016).

[4]Lockdown resulted in lectures and tutorials moving to online at short notice, with the consequent changes in andragogy and assessment.

[5]For a general discussion of misconceptions, see for example Bayraktar (2009), Hammer (1996), Obaidat and Malkawi (2009), and Savinainen and Viiri (2008).

Apart from the previously published works mentioned (Bani-Salameh, 2016a, 2016b; Martín-Blas et al., 2010), there are only a few other studies that explicitly use DMs. For example, in the gender context, Bouzid et al. (2022) has done some work done on Moroccan high school students. Escalante and Cuevas (2023) used the FCI as a pre-entry test for different physics year-levels at a university in Chile, where DMs were then analysed between the year-levels. Although we do not use it here, as were are focused on results for DMs, we should also mention that Martín-Blas et al. (2010) and Escalante and Cuevas (2023) also used 'concentration analysis' (Bao & Redish, 2001) to plot % score versus the concentration of incorrect response for each question.

A related approach to looking at DMs (developed independently) is to look at incorrect responses known as 'polarising choices' (Alinea & Naylor, 2015), where students' answers to particular questions strongly favoured two options: the correct answer and one wrong answer. The authors of those works found that there were subsets of incorrect questions suggestive of a conceptual gap, and as discussed in Carleschi et al. (2022), who found similar issues for the 2020 UJ cohort when compared to the Japanese cohort studied by Alinea and Naylor (2015).

If effective teaching is to take place (Bani-Salameh, 2016a) then a primary objective at the start of teaching, for example classical mechanics, is to identify these misconceptions. Our adaptation of the method of DMs as first discussed by Martín-Blas et al. (2010) is not only to present graphical plots of the correct response percentages, but also to include in the same graph a 'snapshot' of the % of DMs, e.g., see Figure 2 and 3.[6] Given that, like many HEIs, we deploy the FCI through a 'learning management system' (LMS) it is not too difficult to export all the data into a spreadsheet which can be 'processed' to generate such an info-graphic. We shall discuss this approach in the context for first-year undergraduates taking physics at UJ, where in this work, we pursue a multi-year study of the 'pre-test' DMs, displaying a 'snapshot' of 'pre-test' misconceptions over three years (2020 – 2022).

**Research questions**

Given our discussion so far in relation to DMs, we emphasise that we have collected three years of data for a large South African first-year physics course, that can be compared to other countries and also characterised by a large proportion of women. With this in mind, the goal of this work is to explore the following research questions (RQs):

RQ1 – To what extent do DMs change between years for South African first-year physics course students?

RQ2 – How do the DMs in South Africa (at UJ) compare with other countries?

RQ3 – Are there any gender differences in the mean score of the pre-test and the type of DMs?

In what follows we will discuss the methodology, present our results and analyses, and then conclude and suggest recommendations for future work.

**Method**

The recommended means of administering the FCI is as a strictly closed-book test held at the beginning of the semester (the 'pre-test'), to provide instructors with an indication of their students' baseline mechanics skills. The test can be administered on paper, but there are various online versions available, see McKagan et al. (2020) for a discussion of using the PhysPort system in general. Students are not expected to prepare for the assessment, and once the pre-test is completed, it is not reviewed in class nor do students receive feedback on their attempts. At the end of the semester the same FCI test can be administered again (as a 'post-test' to then compute conceptual gains); however, in this article we focus on the 'pre-test' as a way to obtain formative feedback for informed teaching in terms of 'dominant misconceptions' (DMs).

As mentioned earlier we will use the method of DMs as first discussed by Martín-Blas et al. (2010), where we will present for the first time how DMs progress over time for a large cohort of first-year physics students of the order of $N = 400$ for each year (typically $N = 300$ students take the pre-test each year). As far as we are aware, this is the first time that gender has also been considered for DMs at a HEI; see Obaidat and Malkawi (2009) for discussions of DMs and gender in the high school context.

**Participants**

In our present study, we focused on participants who sat the pre-test only and who identified as male or female, leading to a total of $N = 805$ participants, with $N = 130, N = 352$, and $N = 323$ students in each of the 2020−2022 cohorts, respectively. The cohorts consist of Physics Majors, Physics Majors in the Extended Programme, Education, Earth Sciences, Life Sciences, Life Sciences (Ext.) and Engineering students in each consecutive year from 2020 to 2022.[7] We should also mention that the number of participants quoted above is slightly less than the full number of students who sat the pre-test, see Table 1 as an indicative year of our study, because not all participants identified their gender. Note that this year is chosen due to it being separated from the instability caused in previous years from lockdowns. However, the discussed trends in the data were evident in all years. In relation to the fact that we will consider DMs through the 'gender lens', of the $N = 805$ participants in the study in total, there were $N = 333$ women and $N = 472$ men, respectively.

---

[6]Carleschi et al. (2022) and Naylor et al. (2022) used a similar approach with the % correct on the positive axis and the % largest incorrect answer (if greater than the % correct) on the negative axis (not a %nDM).

[7]The 2020 'pre-test' cohort was considerably smaller as this was the only 'pre-test' administered on paper, with testing interrupted due to the onset of the COVID-19 pandemic. In this case, for the Engin. Phys. cohort, as an example, 144 participants out of a possible 404 sat the pre-test in March 2020.

**Table 1**

As an example, course codes for five first-year classes involved in the 2022 FCI pre-test: $N = 337$. Deployment in late February 2022. The notation 28/55 means that 28 participants sat the pre-test from a full cohort of 55.

| Course | Responses |
|---|---|
| PHYS1A1 (Majors) | 28/55 |
| PHY1EA1 (Physics Ext.: Sem1) | 176/250 |
| PHYG1A1 (Earth Sci.) | 9/9 |
| PHYL1A1 (Life Sci.) | 19/40 |
| PHYE0A1 (Engin. Phys.) | 105/500 |

It may also be worth mentioning that the cohort at UJ is an ethnically diverse one including a multitude of cultural and linguistic backgrounds, where in South Africa, the four official population classifications remain African, Coloured, Indian/Asian, and White. This is indicative of the diverse student demographics of the Physics Department at UJ (as the largest first-year module is a service course to the Faculty of Engineering and the Built Environment (FEBE)). Furthermore, there are multiple languages spoken in South Africa (where the country itself has 11 official languages) and at UJ (Faculty of Engineering and the Built Environment [FEBE UJ], 2020)

Further to the interesting ethnic, cultural, and linguistic backgrounds of the students, we are also dealing with large cohorts of students studying a physics course at UJ, of the order of five hundred in each year (a smaller sample, made up of volunteer students, actually sits the 'pre / post-test'). As mentioned earlier, the size of the 2020 cohort who sat the 'pre-test' was smaller than usual due to the onset of the COVID-19 pandemic and the ensuing lockdowns that affected UJ (around March, 2020).

In fact, because of the interruption of the testing process due to the COVID-19 pandemic, we were only able to collect data amounting to about half the size of the other cohorts, cf. Table 1 (for more details see Carleschi et al. (2022)). The 2020 'pre-test' was administered on paper and this does raise some interesting questions, which we will return to in more detail later. However, for all other years, we utilised a learning management system (LMS). The use of a LMS for the deployment of the FCI has several advantages for students and researchers alike: students can write the test in their own time and in a space of their choosing (therefore avoiding the loss of class time), with the aid of assistive technology if required; researchers can deploy the test to a large cohort, grade the tests electronically, as well as collect, store, and process data with greater ease. However, a remaining difficulty lies in capturing paired data i.e. finding student volunteers in statistically significant numbers willing to sit both the pre- and post-test.

Hence, we would argue that a motivation to consider metrics beyond FCI gains, as defined in Eq. (1), is the limitation of the need for the same student to take both the 'pre-' and the 'post-test' (one typically has fewer paired data, even if we have of the order of 300 students take the 'post-test'). As Bani-Salameh has eloquently stated (Bani-Salameh, 2016a), there is a wealth of information that can be gleaned from just the 'pre-test' (or 'post-test') data alone. One of these is the idea of DMs as first discussed by Martín-Blas et al. (2010). From this idea, we can not only investigate how many students correctly answered a particular question (in the FCI each question forms part of several concepts in classical mechanics, see Tables A1 and A2), but we can also ask if there were any particular questions that had dominant incorrect responses (Bani-Salameh, 2016b; Martín-Blas et al., 2010). We will discuss DMs in more detail in the next section.

**Data analysis**

While we could use methods like IRT or the Rasch model, e.g., see Morris et al. (2006) and Nitta and Aiba (2019), to investigate how questions and students fare in terms of difficulty and ability, we have in mind the usage of a relatively fast and still visual approach to seeing how a given class or cohort performs at the pre-test stage of the FCI. The method of DMs can still be used to compare a given cohort using a 'post-' vs. 'pre-test' and we leave this comparison for future work.

**Normalised dominant misconceptions**

To help explain what is meant by a 'normalised' DM (nDM), as in Martín-Blas et al. (2010), we will assume there are 5 options $-$ $A, B, C, D, E$ $-$ and the option of not answering at all, $N/A$, see our earlier discussion in relation to Figure 1. The following equation can be used to calculate the %nDM for a given question,

$$\%nDM = \frac{\%(\text{largest incorrect answer})}{100 - \%(\text{correct answer}) - \%(N/A)} \times 100 \ . \quad (3)$$

The DMs are normalised by subtracting away the %(correct answer) and a threshold is chosen where anything above this 'value' is deemed a DM. For example, this means that in many cases some question items do not have a DM (i.e. answers lie below a certain % threshold), or that there could be more than one incorrect choice which is dominant. As further discussed by Bani-Salameh (2016a), the actual value for the threshold (the % where the question is defined as dominant) can be moved. While Martín-Blas et al. (2010) chose 50% for their threshold, we have decided to choose 60% for the UJ cohorts.

Our choice of normalisation in Eq. (3) relates to the issue of 'student apathy' in relation to students who skipped questions or who quit the test before they completed a response. We have chosen to include all students who 'attempted' the 'pre-test'. Usually when considering the normalised gain (see Eq. (1)), the proviso of students answering over 20% of all questions on the FCI is often stated (Fazio & Battaglia, 2019) and is the reason why many authors remove (clean) data for those students who have given a 'no-answer' or 'N/A' in a certain % of questions. However, for DMs we can include such cases as we are able to 'normalise' the DMs (see Eq. (3)) such that we can account for students who did not answer. We are of the view that Eq. (3) is a better representation of DMs to compare the relative size of the largest incorrect answer as compared to all the other incorrect answers by subtracting away the %(correct answer) and the %N/A. [8]

**Table 2**

Selected Newtonian concepts in the force concept inventory (FCI).

| Qn. | Newtonian Concept[a] |
|---|---|
| 4, 28 | Non-equal action-reaction pairs |
| 19 | Position/velocity/acceleration undiscr. |
| 2 | Displacement time depends on the mass |
| 9 | Non-vectorial velocity composition |
| 14 | Ego-centred reference frame |

[a]The Newtonian concepts considered here come from the taxonomy described in Tables A1 and A2. Note undiscr. is short for undiscriminated

**Comparison with other countries and gender**

A further benefit of using DMs is that we can compare this with other groups of students outside of South Africa, such as those in Spain (Martín-Blas et al., 2010) and the Kingdom of Saudi Arabia (KSA) (Bani-Salameh, 2016a, 2016b), and as we mentioned for gender for the UJ students. Complete tables of data for all questions can be found in those works, respectively. However, Martín-Blas et al. (2010) found a given set of questions that had large DMs and were well correlated between the two cohorts studied (Martín-Blas et al., 2010) and relate to particular Newtonian concepts – these are presented in Table 2. These particular concepts have been broadly classified based on the taxonomy of FCI questions as first developed by Hestenes et al. (1992) and adapted by others (Bani-Salameh, 2016a; Bayraktar, 2009; Martín-Blas et al., 2010). We present them here, for completeness, with our own slight adaptations in Appendix A (supplemental); see Tables A1 and A2. The is the reason for the question choices and ordering given in Tables 4 and 6 in the following sections.

**Results**

**Three-years of dominant misconceptions**

In this section we first present our year-by-year comparison of data (from 2020 – 2022), see Figure 2, based on $N = 130$, $N = 352$ and $N = 323$ students, respectively. The total over three years, $N = 805$, is presented in Figure 3, along with results for gender (to be discussed later). The positive vertical axis corresponds to the % correct answer for each question number. The negative vertical axis shows the %nDM for each question.

In these figures we have removed the question numbers which have a %nDM scores of less than 50% – these are to be classified as non-DMs and below threshold. This is inline with the thresholds set by Bani-Salameh (2016a) and Martín-Blas et al. (2010). For further clarity, leading to a smaller set of DMs, we have imposed an upper DM threshold for all cohorts at 60%, the bold horizontal lines in Figures 2 and 3.

In Figure 2 the key can be used for each of the 2020, 2021, and 2022 years, respectively. What is particularly useful with these representations is that for a given class/cohort, a 'snapshot' of the DMs (as well as the % correct answers) can be obtained. Different class cohorts and year levels will typically show slightly different types of DMs. For example, when comparing the number of DMs for the 2020 cohort as compared to the other years (2021, 2022), we can clearly see that there are more DMs (above the 60% threshold) and more questions highlighted at above 50%. One might also be tempted to infer that the 2020 cohort did better on the % correct answers, based on the means scores presented in the Table 3.

**Table 3**

The mean ($\bar{x}$) for (%correct answer)s and %nDMs for the yearly cohorts and full three-years, with participant size $N$; standard deviation in brackets: (SD).

| Year | $N$ | $\bar{x}_{\%Corr}$ (SD) | $\bar{x}_{\%nDM}$ (SD) |
|---|---|---|---|
| 2020 | 130 | 34.3 (14.7) | 59.6 (13.8) |
| 2021 | 352 | 30.9 (13.5) | 49.4 (14.0) |
| 2022 | 323 | 30.1 (13.1) | 50.6 (13.9) |
| 2020-2022 | 805 | 31.2 (13.2) | 51.7 (13.8) |

We can investigate these two issues further by looking at Table 3, where the 2021 and 2022 cohorts appear to have performed better, as suggested by their smaller mean %nDMs

---

[8]The %NA normalisation effect is very small in practice as less than 1% of all students in the study attempted less than 20% of all questions: less than 10 students from $N = 805$.

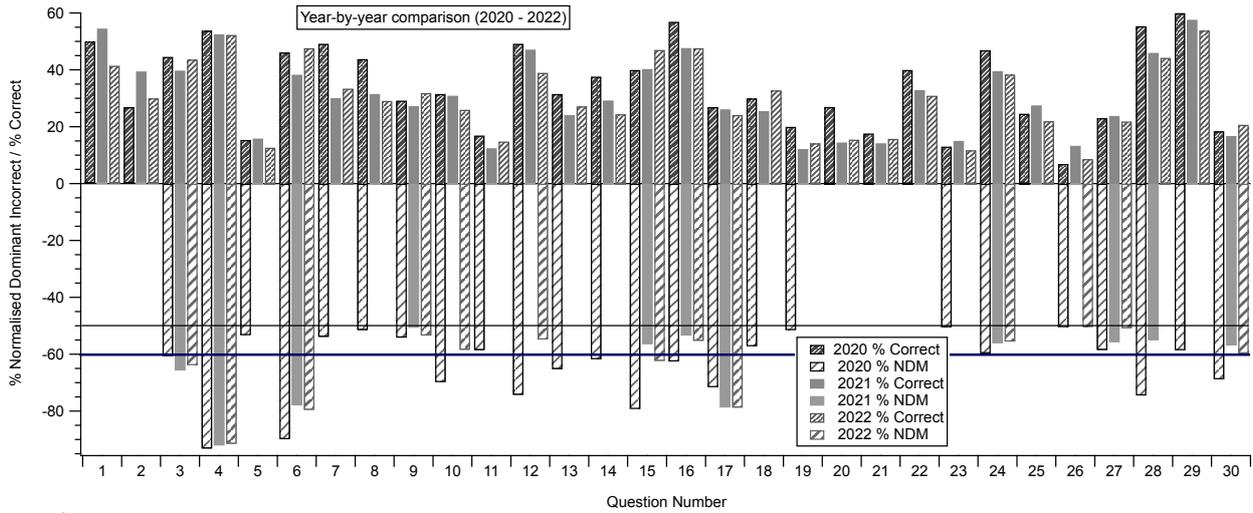

**Figure 2**

Pre-test year-by-year breakdown for the 2020 − 2022 cohorts, sitting an introductory physics course, $N = 130, 352, 323$. For clarity we have removed DM question items with a %nDM score less than 50%, horizontal line at $y = −50$%. We have set the upper threshold to $y = −60$% (bold horizontal line).

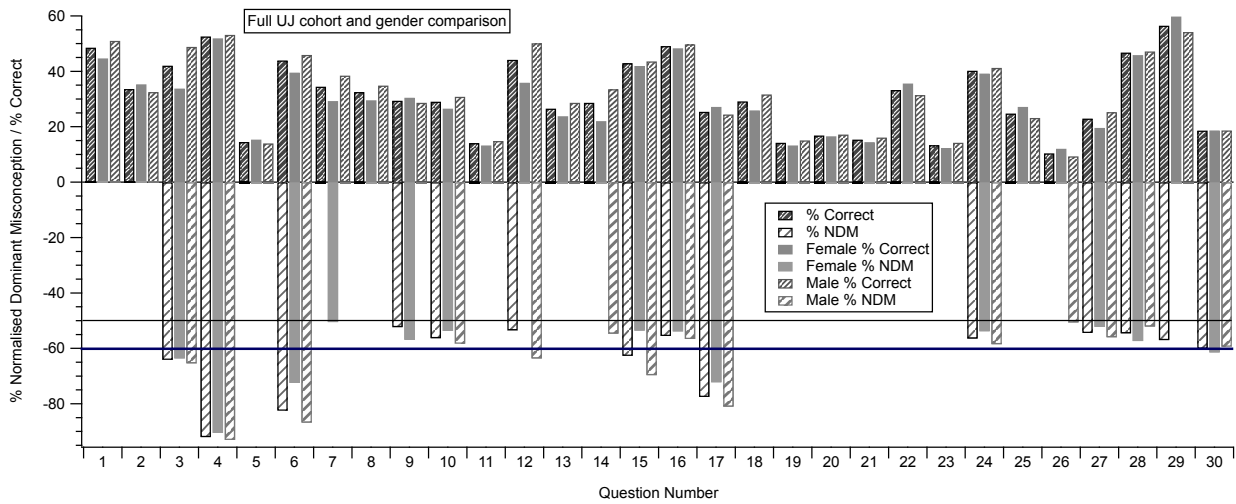

**Figure 3**

Pre-test scores for the full no. of students, $N = 805$ (over three years, from 2020 − 2022), compared to the same cohort broken down into participants who identify as female, $N = 333$, and male, $N = 472$. Same threshold: $y = −60$%.

when compared against the 2020 cohort; this trend is not observed in the % correct answers.[9] To verify this, a 'one-way analysis of means' (i.e. not assuming equal variances, see Table 3) (R Core Team, 2022) shows that at the 95% confidence level, the difference in the means of DMs, when comparing 2020 to the 2021 and 2022 cohorts, is statistically significant: $F = 4.6522$ and p-value = 0.01337 (Welch's test). The difference in means for the % correct answers was not found to be statistically significant and satisfied the null hypothesis (cf. Figure 2).

**Comparison with other countries**

As a further example, we have also presented comparisons of our results at UJ with data from other countries in Table 4. We used the questions discussed by Martín-Blas et al. (2010) (who considered two groups of Spanish engineering students) along with data from the KSA (Bani-Salameh,

---

[9]Note, the average of the summed mean scores will not be equal to the mean of all summed scores together, because the participant groups sizes are different for each year, see Table 3.

**Table 4**

A comparison of %nDMs from Martín-Blas et al. (2010) (groups 1 and 2) and Bani-Salameh (2016b) (rounded to nearest whole number) and UJ cohorts from 2020 to 2022. Underlined scores for the average for UJ, for Q2 and $Q$14 show a significant difference (compared to the half-range of the 2020 − 2022 means) in the associated Newtonian concepts, see Table 2.

|  | Q4 (%) | Q28 (%) | Q19 (%) | Q2 (%) | Q9 (%) | Q14 (%) |
|---|---|---|---|---|---|---|
| Group 1 (MB) | 95 | 68 | 44 | 68 | 56 | 96 |
| Group 2 (MB) | 96 | 64 | 58 | 64 | 65 | 86 |
| Bani-Salameh | 73 | 56 | 67 | 58 | 49 | 90 |
| UJ 2020 | 93 | 74 | 51 | 35 | 54 | 62 |
| UJ 2021 | 92 | 55 | 46 | 32 | 51 | 43 |
| UJ 2022 | 91 | 46 | 40 | 30 | 53 | 48 |
| UJ (2020-2022) | 92 | 54 | 44 | <u>32</u> | 52 | <u>48</u> |

2016a) to compare with UJ students. In Table 4 we can see similarities between the DMs of Martín-Blas et al. (Group 1 and Group 2), Bani-Salameh and with the UJ cohorts.

However, from looking at the UJ cohort over three years from 2020 to 2022, we notice differences as compared to that of Martín-Blas et al. (2010) and Bani-Salameh (2016a). The underlined questions in Table 4 for the average over three years at UJ (for $Q$2 and $Q$14), show a significant departure (determined from half the range of the UJ, 2020 − 2022 means) in the associated Newtonian concepts as compared to the Spanish cohorts. From Table 2 it can be seen that these two questions correspond to displacement time depending on mass and ego-centred reference frames.

**Table 5**

The mean ($\bar{x}$) % of correct answers and %nDMs; standard deviation in brackets: (SD).

| Gender | N | $\bar{x}_{\%Corr}$ (SD) | $\bar{x}_{\%nDM}$ (SD) |
|---|---|---|---|
| Male | 472 | 32.2 (13.7) | 53.0 (14.7) |
| Female | 333 | 29.7 (13.0) | 48.0 (14.5) |

**DMs through the gender lens**

As mentioned in the *Introduction* the performance difference between gender has been well discussed (Docktor & Heller, 2008; Madsen et al., 2013) for the FCI. As far as we are aware, this is the first time that DMs have been considered through a 'gender lens', apart from research on Moroccan high school students (Bouzid et al., 2022), rather than at a HEI. The gender results are presented graphically in Figure 3, where in particular there are differences in the question type for DMs.

For example the male cohort has $Q$12 as a DM while the female cohort does not. We can then refer to the taxonomy in Table A2 to see that $Q$12 relates to projectile motion, also see Figure 1. In the case of this question in both cohorts the DM choice was response *C* (see Table B1) which is CI1 from Table A1: 'largest force determines motion'. Further questions can be analysed in a similar manner, particularly in relation to DMs. For clarity we have presented a full comparison of the differences in DM choice between genders in Appendix B.

Like we did for the year-by-year data we have presented a comparison between gender for the means of the %(correct answer)s and %nDMs in Table 5. Interestingly, up to the fact that the % differences are small, the male cohort has a slightly higher mean % correct score; in the same instance the female cohort has a lower %nDM. However, we have checked using the 'independent samples t-test' (R Core Team, 2022) to verify that at the 95% confidence level the differences are not significant and satisfy the null hypothesis. Our data suggests then that there is no statistically significant difference in 'pre-test' FCI performance of students who identity as male and female participating in our study.

Finally, in Table 6 a country comparison of UJ gender and the full Spanish and the KSA cohorts has been made based on the questions chosen by Martín-Blas et al. (2010). Specifically, we divide the data collected over the three-year period (2020 − 2022) from the $N$ = 805 total students into two categories: male participants ($N$ = 472) and female participants ($N$ = 333). Similar % scores are seen for several questions, with the exception of lower scores for $Q$2, $Q$14, and − to a lesser extent − $Q$9. A more detailed analysis of these 'gender' issues is left for future work.

**Discussion**

The results from the previous section have highlighted how the idea of 'dominant misconceptions' (DMs) can be used in the FCI at the 'pre-test' stage as 'assessment for learning' (Gioka, 2006) where figures such as those presented in Figures 2 and 3 allows the teacher / lecturer to have a 'snapshot' of a given cohort's conceptual understanding and adapt and respond to the students' learning needs. For example, in the gender context, we have noted that both men and women in relation to $Q$12, see Figure 1, had the same DM: 'largest force determines motion'. A lecturer aware of

**Table 6**

A comparison of %DMs from Martín-Blas et al., 2010 (groups 1 and 2, Spain) and Bani-Salameh, 2016b (KSA) rounded to nearest whole number, with gender on the UJ cohorts. Underlined notation is the same as in Table 4.

|  | Q4 (%) | Q28 (%) | Q19 (%) | Q2 (%) | Q9 (%) | Q14 (%) |
|---|---|---|---|---|---|---|
| Group 1 (MB) | 95 | 68 | 44 | 68 | 56 | 96 |
| Group 2 (MB) | 96 | 64 | 58 | 64 | 65 | 86 |
| Bani-Salameh | 73 | 56 | 67 | 58 | 49 | 90 |
| UJ Male | 93 | 52 | 46 | <u>33</u> | 49 | <u>55</u> |
| UJ Female | 90 | 57 | 43 | <u>29</u> | 57 | <u>43</u> |

this misconception at the onset of a course can then save additional time and resources in demonstrating that this DM is not so through creative counterexamples, across both the female/male cohort.

This is supported by previous research on DMs by Bani-Salameh (2016a, 2016b), Bouzid et al. (2022), and Martín-Blas et al. (2010) that has elegantly highlighted how DMs can be employed to adapt one's teaching. These works tend to favour systematic tabular presentations spread over several pages. However, we choose instead a more graphical approach that allows us to condense a large volume of data (which may be useful in terms of a suggested policy for HEIs). We present the %(correct answer) on the positive vertical axis with the %nDM on the negative axis (with thresholds), see Figure 2 for year-by-year comparisons; and cf. Figure 3 for a total and gender comparison. This 'dashboard' format allows the lecturer to 'absorb and adapt' to the conceptual issues of each given class.

In response to our initial research questions, we have found that:

> RQ1 − There was a statistically significant % mean score difference for DMs that was larger for 2020 as compared to the 2021 and 2022 first-year physics students, see Table 3.
>
> RQ2 − There were only significant differences in two of the Newtonian concepts presented in Table 2 as compared to other countries: *Q*2 and *Q*14 which relate to displacement time and mass, and ego-centred references frames, respectively, see Table 4 and 6.
>
> RQ3 − There were no statistically significant differences in the % mean score of DMs between gender for students taken from 2020 − 2022, see Table 5. However, there were some differences between the choice of dominant incorrect response, see Table B1.

In relation to RQ1, large scale disruptions in 2020 (due to COVID-19) may have been the reason for this effect, where the 2020 cohort had a higher mean %nDM score than the 2021, 2022 cohorts.[10] This has possible connections with work of Lee et al. (2021) on rapid feedback and anxiety, where the 2020 cohort were likely to have had higher than usual levels of anxiety. We highlight also that the 2020 cohort was the only cohort who wrote the FCI test in person, in a test venue on the UJ campus, which may have exacerbated this. This issue requires further study.

There also appears to be an interesting trend in the way the questions are answered in that the %N/A questions increases towards the end of the test. This relates in general to what we defined earlier as a form of 'student apathy.' However, this may not be the whole story; many respondents indicated in a survey (forthcoming work) that they required more time to complete the FCI. This may be particularly so for students at UJ and in South Africa, in which many students do not have English as their first language, and therefore expend extra time and energy translating to their home language.

**Conclusion**

In summary, this work has discussed a three-year comparative study of the DMs, at the pre-test stage, for first-year physics students at UJ, South Africa. We looked at year-to-year changes (from 2020 − 2022) finding some differences, particularly for 2020, which may have been due to COVID-19, as alluded to earlier. The UJ first-year physics students were also compared to other countries, where there were some difference in the type of DM observed. Finally, we also found no significant difference in the mean % scores for DMs between gender.

It is worth emphasising that this graphical representation of the method of DMs is well suited to large cohorts where various LMSs could be used to collect data that is then exported as a 'dashboard' info-graphic. This can be employed at the 'pre-test' and optionally at the 'post-test' phase (no paired data is needed unlike for the Hake Gain, cf Eq. (1)). The 'pre-test' infographical method could even be used to track years levels as in the work of Escalante and Cuevas (2023). As we discussed in the *Introduction*, it is also important to consider the gains in a more mathematically complete way using Rasch Gain (Nitta & Aiba, 2019) or IRT methods (Wang & Bao, 2010), and in the UJ context this can be done with several years' worth of data, including gender.

---

[10]The statistical analyses involved a one-way ANOVA of multiple groups (RQ1) and the student t-test (RQ2), at the 95% confidence level (R Core Team, 2022).

As we mentioned earlier, the multicultural nature of the UJ cohorts (FEBE UJ, 2020) does have implications for the use of the FCI. For the case of UJ, the FCI was deployed using the English version – currently there are no versions of the FCI available in African languages (McKagan et al., 2020). Giving students the opportunity to write the FCI in their preferred language, such that they do not have to expend unnecessary time and energy to understand the test questions, may affect the FCI completion time and overall scores.

Following the ideas mentioned in Hewagallage et al. (2022), it would be interesting to use tools such as 'structured equation modelling' (SEM) (Cwik & Singh, 2022) to look at high school matriculation results and their correlations to 'pre-test' scores at UJ. What may be significant, in the South African context, is whether or not the scores of South African students on their English matriculation results (and correlations with the FCI) depend on their culturally diverse language backgrounds. In this vein, a study on the correlations between students' matriculation results for physical science and/or mathematics with their FCI results may also provide insights on how well students are prepared by their secondary education for university-level physics courses.

In conclusion, we have discussed how the method of DMs can be portrayed graphically to give a 'snapshot' of a physics cohort's 'misconceptions' that allows for the FCI to be *efficiently* used as 'assessment for learning'. This is rather than having to wait for 'post-test' data to work out gains or employing other labour intensive statistical analyses, such as IRT. This enables a lecturer/teacher to make evidence-based decisions for their teaching on a classical mechanics course, for example. As future work, it would also be interesting to look at how 'post-test' DMs compare (Bani-Salameh, 2016b) between year cohorts and gender at UJ.

## Acknowledgements

AC is supported by the NRF and Department of Science and Innovation through the SA-CERN programme, as well as by a Campus France scholarship. ASC is supported in part by the National Research Foundation of South Africa (NRF). WN is supported by a Faculty of Education and Arts Grant: Project Code No: 50-905300-111. The authors express their sincere thanks to the participating students, as well as to the lecturers in the Physics Department and Blackboard technical support at UJ who authorised and facilitated access to the above-mentioned course modules. In particular we are indebted to Emanuela Carleschi for support on this project and for the many warm and stimulating conversations that led to this work.

## Ethical statement

The authors confirm that all participants in the study gave consent to participate in the study, and consent for the results to be published was obtained from participants for all research involving human subjects. The request to conduct this study (in accordance with the principles embodied in the Declaration of Helsinki and local statutory requirements) for the project entitled 'Identifying conceptual issues in first year physics curricula,' was granted by the ethics committee at the University of Johannesburg (UJ): Project Code No: 2023-02-03/Cornell-R1. This includes studies done on teaching methods.

## References

Alinea, A. L., & Naylor, W. (2015). Polarization of physics on global courses. *Physics Education*, *50*, 210–217. https://doi.org/https://doi.org/10.1088/0031-9120/50/2/210

Alinea, A. L., & Naylor, W. (2017). Gender gap and polarisation of physics on global courses. *Physics Education, IAPT*, *33*. http://www.physedu.in/pub/Apr-Jun-2017/PE17-02-425

Australian Curriculum and Assessment Reporting Authority. (2017). National literacy and numeracy learning progressions. https://www.australiancurriculum.edu.au/resources/national-literacy-and-numeracy-learning-progressions/

Bandura, A. (1977). Self-efficacy: Toward a unifying theory of behavioral change. *Psychological review*, *84*(2), 191.

Bani-Salameh, H. N. (2016a). How persistent are the misconceptions about force and motion held by college students? *Physics Education*, *52*(1), 014003. https://doi.org/10.1088/1361-6552/52/1/014003

Bani-Salameh, H. N. (2016b). Using the method of dominant incorrect answers with the FCI test to diagnose misconceptions held by first year college students. *Physics Education*, *52*(1), 015006. https://doi.org/10.1088/1361-6552/52/1/015006

Bao, L., & Redish, E. F. (2001). Concentration analysis: A quantitative assessment of student states. *American Journal of Physics*, *69*(S1), S45–S53. https://doi.org/10.1119/1.1371253

Bates, S., Donnelly, R., Macphee, C., Sands, D., Birch, M., & Walet, N. (2013). Gender differences in conceptual understanding of newtonian mechanics: A uk cross-institution comparison. *European Journal of Physics*, *34*(2), 421–434. https://doi.org/10.1088/0143-0807/34/2/421

Bayraktar, S. (2009). Misconceptions of turkish pre-service teachers about force and motion. *International*

*Journal of Science and Mathematics Education*, *7*(2), 273–291.

Black, P., & Wiliam, D. (1998). Assessment and classroom learning. *Assessment in Education: Principles, Policy & Practice*, *5*(1), 7–74. https://doi.org/10.1080/0969595980050102

Bouzid, T., Kaddari, F., & Darhmaoui, H. (2022). Force and motion misconceptions' pliability, the case of moroccan high school students. *The Journal of Educational Research*, *115*(2), 122–132. https://doi.org/10.1080/00220671.2022.2064802

Carleschi, E., Chrysostomou, A., Cornell, A. S., & Naylor, W. (2022). Probing the effect on student conceptual understanding due to a forced mid-semester transition to online teaching. *European Journal of Physics*, *43*(3), 035702. https://doi.org/10.1088/1361-6404/ac41d9

Clement, J., Brown, D. E., & Zietsman, A. (1989). Not all preconceptions are misconceptions: Finding 'anchoring conceptions' for grounding instruction on students' intuitions. *International Journal of Science Education*, *11*(5), 554–565. https://doi.org/10.1080/0950069890110507

Coletta, V. (2013). Reducing the fci gender gap. *Physics Education Research Conference 2013*, 101–104.

Cwik, S., & Singh, C. (2022). Gender differences in students' self-efficacy in introductory physics courses in which women outnumber men predict their grade. *Phys. Rev. Phys. Educ. Res.*, *18*, 020142. https://doi.org/10.1103/PhysRevPhysEducRes.18.020142

Docktor, J., & Heller, K. (2008). Gender differences in both force concept inventory and introductory physics performance. *AIP Conference Proceedings*, *1064*(1), 15–18. https://doi.org/10.1063/1.3021243

Eddy, S. L., & Brownell, S. E. (2016). Beneath the numbers: A review of gender disparities in undergraduate education across science, technology, engineering, and math disciplines. *Phys. Rev. Phys. Educ. Res.*, *12*, 020106. https://doi.org/10.1103/PhysRevPhysEducRes.12.020106

Escalante, F., & Cuevas, F. (2023). Comparing the acquisition of concepts in Newtonian mechanics for engineering students in different levels courses [050002]. *AIP Conference Proceedings*, *2731*(1). https://doi.org/10.1063/5.0133083

Faculty of Engineering and the Built Environment. (2020). University of Johannesburg Faculty of Engineering and the Built Environment 2019 Annual Report [https://www.uj.ac.za/faculties/febe/Pages/annual-report.aspx].

Fagen, A. P., Crouch, C. H., & Mazur, E. (2002). Peer instruction: Results from a range of classrooms. *The Physics Teacher*, *40*(4), 206–209. https://doi.org/10.1119/1.1474140

Fazio, C., & Battaglia, O. R. (2019). Conceptual understanding of newtonian mechanics through cluster analysis of fci student answers. *International Journal of Science and Mathematics Education*, *17*(8), 1497–1517.

Francis, D. A., Brown, A., McAllister, J., Mosime, S. T., Thani, G. T. Q., Reygan, F., Dlamini, B., Nogela, L., & Muller, M. (2019). A five country study of gender and sexuality diversity and schooling in southern africa. *Africa Education Review*, *16*(1), 19–39. https://doi.org/10.1080/18146627.2017.1359637

Fulmer, G. W., Liang, L. L., & Liu, X. (2014). Applying a force and motion learning progression over an extended time span using the force concept inventory. *International Journal of Science Education*, *36*(17), 2918–2936. https://doi.org/10.1080/09500693.2014.939120

Gioka, O. (2006). Assessment for learning in physics investigations: Assessment criteria, questions and feedback in marking. *Physics Education*, *41*(4), 342–346. https://doi.org/https://doi.org/10.1088/0031-9120/41/4/009

Glasser, H. M., & John P. Smith, I. (2008). On the vague meaning of "gender" in education research: The problem, its sources, and recommendations for practice. *Educational Researcher*, *37*(6), 343–350. https://doi.org/10.3102/0013189X08323718

Hake, R. R. (1998). Interactive-engagement versus traditional methods: A six-thousand-student survey of mechanics test data for introductory physics courses. *American Journal of Physics*, *66*(1), 64–74. https://doi.org/10.1119/1.18809

Halloun, I. A., & Hestenes, D. (1985). The initial knowledge state of college physics students. *American Journal of Physics*, *53*, 1043.

Hammer, D. (1996). More than misconceptions: Multiple perspectives on student knowledge and reasoning, and an appropriate role for education research. *American Journal of Physics*, *64*(10), 1316–1325. https://doi.org/10.1119/1.18376

Hestenes, D. (1998). Who needs physics education research!? *American Journal of Physics*, *66*, 465.

Hestenes, D., Wells, M., & Swackhamer, G. (1992). Force concept inventory. *The Physics Teacher*, *30*(3), 141–158. https://doi.org/10.1119/1.2343497

Hewagallage, D., Christman, E., & Stewart, J. (2022). Examining the relation of high school preparation and college achievement to conceptual understanding. *Phys. Rev. Phys. Educ. Res.*, *18*, 010149. https://doi.org/10.1103/PhysRevPhysEducRes.18.010149

Lee, S., Choi, Y.-i., & Kim, S.-W. (2021). Roles of emotions induced by immediate feedback in a physics problem-solving activity. *International Journal of Science Education*, *43*(10), 1525–1553. https://doi.org/10.1080/09500693.2021.1922778

Madsen, A., McKagan, S. B., & Sayre, E. C. (2013). Gender gap on concept inventories in physics: What is consistent, what is inconsistent, and what factors influence the gap? *Phys. Rev. ST Phys. Educ. Res.*, *9*, 020121. https://doi.org/10.1103/PhysRevSTPER.9.020121

Madsen, A., McKagan, S. B., & Sayre, E. C. (2017). Best practices for administering concept inventories. *The Physics Teacher*, *55*(9), 530–536. https://doi.org/10.1119/1.5011826

Main, P. (2022). *Assessment in university physics education*. IOP Publishing. https://iopscience.iop.org/book/mono/978-0-7503-3851-6

Malespina, A., & Singh, C. (2022). Gender differences in test anxiety and self-efficacy: Why instructors should emphasize low-stakes formative assessments in physics courses. *European Journal of Physics*, *43*(3), 035701. https://doi.org/10.1088/1361-6404/ac51b1

Marchand, G. C., & Taasoobshirazi, G. (2013). Stereotype threat and women's performance in physics. *International Journal of Science Education*, *35*(18), 3050–3061. https://doi.org/10.1080/09500693.2012.683461

Maries, A., & Singh, C. (2016). Teaching assistants' performance at identifying common introductory student difficulties in mechanics revealed by the force concept inventory. *Phys. Rev. Phys. Educ. Res.*, *12*, 010131. https://doi.org/10.1103/PhysRevPhysEducRes.12.010131

Marshman, E. M., Kalender, Z. Y., Nokes-Malach, T., Schunn, C., & Singh, C. (2018). Female students with a's have similar physics self-efficacy as male students with c's in introductory courses: A cause for alarm? *Phys. Rev. Phys. Educ. Res.*, *14*, 020123. https://doi.org/10.1103/PhysRevPhysEducRes.14.020123

Martín-Blas, T., Seidel, L., & Serrano-Fernández, A. (2010). Enhancing force concept inventory diagnostics to identify dominant misconceptions in first-year engineering physics. *European Journal of Engineering Education*, *35*(6), 597–606. https://doi.org/10.1080/03043797.2010.497552

Mazur, E. (1997). *Peer instruction: A user's manual*. Prentice Hall. /files/mazur/files/rep_0.pdf

McKagan, S. B., Strubbe, L. E., Barbato, L. J., Mason, B. A., Madsen, A. M., & Sayre, E. C. (2020). PhysPort use and growth: Supporting physics teaching with research-based resources since 2011. *The Physics Teacher*, *58*(7), 465–469. https://doi.org/10.1119/10.0002062

Morris, G. A., Branum-Martin, L., Harshman, N., Baker, S. D., Mazur, E., Dutta, S., Mzoughi, T., & McCauley, V. (2006). Testing the test: Item response curves and test quality. *American Journal of Physics*, *74*(5), 449–453. https://doi.org/10.1119/1.2174053

National Council of Teachers of Mathematics. (1995). *Assessment standards for school mathematics*.

National Research Council. (2012). *A framework for k-12 science education: Practices, crosscutting concepts, and core ideas*. The National Academies Press.

Naylor, W., Chrysostomou, A., Carleschi, E., & Cornell, A. S. (2022). Using technology to understand student 'misconceptions' in classical mechanics. In K. Menon, K. Naidoo, & G. Castrillon (Eds.). Jacana Media. https://www.mydigitalpublication.co.za/uj/teaching-innovation-dte/data/mobile/index.html

Nissen, J. M., Talbot, R. M., Nasim Thompson, A., & Van Dusen, B. (2018). Comparison of normalized gain and cohen's d for analyzing gains on concept inventories. *Phys. Rev. Phys. Educ. Res.*, *14*, 010115. https://doi.org/10.1103/PhysRevPhysEducRes.14.010115

Nitta, H., & Aiba, T. (2019). An alternative learning gain based on the rasch model. *The Physics Educator*, *01*(01), 1950005. https://doi.org/10.1142/S2661339519500057

Obaidat, I., & Malkawi, E. (2009). The grasp of physics concepts of motion: Identifying particular patterns in students' thinking. *International Journal for the Scholarship of Teaching and Learning*, *3*(1), n1.

Planinic, M., Ivanjek, L., & Susac, A. (2010). Rasch model based analysis of the force concept inventory. *Phys. Rev. ST Phys. Educ. Res.*, *6*, 010103. https://doi.org/10.1103/PhysRevSTPER.6.010103

R Core Team. (2022). *R: A language and environment for statistical computing*. R Foundation for Statistical Computing. Vienna, Austria. http://www.R-project.org/

Ryan, R. M., & Deci, E. L. (2000). Self-determination theory and the facilitation of intrinsic motivation, social development, and well-being. *American Psychologist*, *55*, 68–78. https://doi.org/https://psycnet.apa.org/doi/10.1037/0003-066X.55.1.68

Sá, C., Cowley, S., Martinez, M., Kachynska, N., & Sabzalieva, E. (2020). Gender gaps in research productivity and recognition among elite scientists in the U.S., canada, and south africa. *PLoS One*, *15*(10), e0240903.

Savinainen, A., & Viiri, J. (2008). The force concept inventory as a measure of students conceptual coherence. *International Journal of Science and Mathematics Education*, 6(4), 719–740.

Shapiro, J. R., & Williams, A. (2012). The role of stereotype threats in undermining girls' and women's performance and interest in stem fields. *Sex Roles*, 66, 175–183.

Stoen, S. M., McDaniel, M. A., Frey, R. F., Hynes, K. M., & Cahill, M. J. (2020). Force concept inventory: More than just conceptual understanding. *Phys. Rev. Phys. Educ. Res.*, 16, 010105. https://doi.org/10.1103/PhysRevPhysEducRes.16.010105

Wang, J., & Bao, L. (2010). Analyzing force concept inventory with item response theory. *American Journal of Physics*, 78(10), 1064–1070. https://doi.org/10.1119/1.3443565

Yasuda, J., Mae, N., Hull, M. M., & Taniguchi, M. (2021). Analysis to develop computerized adaptive testing with the force concept inventory. *1929*(1), 012009. https://doi.org/10.1088/1742-6596/1929/1/012009

Zimmerman, B. J. (2000). Self-efficacy: An essential motive to learn. *Contemporary Educational Psychology*, 25(1), 82–91. https://doi.org/https://doi.org/10.1006/ceps.1999.1016

## Appendix A
**Supplemental: Taxonomy of Concepts for FCI Questions**

For completeness, we present two tables to help describe the taxonomy of conceptual understanding and hence possible 'misconception' issues for each question in the FCI. These tables are based on the works of Bani-Salameh (2016b), Bayraktar (2009), and Hestenes et al. (1992). The table below, Table A1, is used for the conceptual breakdown of FCI questions used in Table A2.

### Table A1

Conceptual coding for the FCI; based on Bani-Salameh (2016b).

| Code | Misconception (Pre-conception) |
|------|-------------------------------|
| K1   | Position-velocity undiscriminated |
| K2   | Velocity-acceleration undiscriminated |
| K3   | Non-vectorial velocity composition |
| K4   | Ego-centered reference frame |
| I1   | Impetus supplied by 'hit' |
| I2   | Loss/recovery of original impetus |
| I3   | Impetus dissipation |
| I4   | Gradual/delayed impetus build-up |
| I5   | Circular impetus |
| AF1  | Only active agents exert forces |
| AF2  | Motion implies active force |
| AF3  | No motion implies no force |
| AF4  | Velocity proportional to applied force |
| AF5  | Acceleration implies increasing force |
| AF6  | Force causes acceleration to terminal velocity |
| AF7  | Active force wears out |
| AR1  | Greater mass implies greater force |
| AR2  | Most active agent produces greatest force |
| CI1  | Largest force determines motion |
| CI2  | Force compromise determines motion |
| CI3  | Last force to act determines motion |
| CF   | Centrifugal force |
| Ob   | Obstacles exert no force |
| R1   | Mass makes things stop |
| R2   | Motion when force overcomes resistance |
| R3   | Resistance opposes force/impetus |
| G1   | Air pressure-assisted gravity |
| G2   | Gravity intrinsic to mass |
| G3   | Heavier objects fall faster |
| G4   | Gravity increases as objects fall |
| G5   | Gravity acts after impetus wears down |

Using Table A1 above, the following table, Table A2, gives information about the breakdown of each question response conceptually and is similar in vein to that presented in the work of Bani-Salameh (2016a). Note that not every answer response (A, B, C, D, E) has a conceptual code.

## Appendix B
**DM choice and gender**

We can also make a comparison of the largest incorrect choice (DM) between the female and male cohorts (from

**Table A2**

Conceptual breakdown of FCI questions 1 − 30. These concepts are grouped roughly into: Newton's three laws, circular motion, s vs t graphs, projectile motion, vector addition and $\Sigma F_{net} = 0$. The coding of each question response (from A to E) is given in Table A1. Adapted from Bani-Salameh (2016b).

|    | Law/concept           | A        | B         | C         | D          | E              |
|----|-----------------------|----------|-----------|-----------|------------|----------------|
| 1  | Newton II             | G3       |           |           |            |                |
| 2  | Newton II             |          | G3        |           | G3         |                |
| 3  | Newton II             | AF6      | AF5, G4   |           | G2         | G1             |
| 4  | Newton II, III        | AR1      |           | Ob        | AR1        |                |
| 5  | circ. motion          | Ob       |           | I1, I5, AF2 | I1, I5, AF2 | I1, I5, AF2, CF |
| 6  | circ. motion          | I5       |           | CF        | CI2, CF    | CF             |
| 7  | circ. motion          | I5       |           | CI2, CF   | I2, I5, CF | CF             |
| 8  | s vs t                | CI3      |           | I2        | I4         | I2             |
| 9  | vec. add.             |          | CI3       | K3        |            |                |
| 10 | $\Sigma F_{net} = 0$  |          | I4        |           | I4         |                |
| 11 | $\Sigma F_{net} = 0$  | Ob, G1   | I1, Ob    | I1        |            | G2             |
| 12 | proj. motion          | CI2      |           | I3        | I3, G5     |                |
| 13 | Newton II             | I3       | I3, G4, G5| I3        |            | G2             |
| 14 | proj. motion          | K4       | K4        | CI2       |            | I3, G5         |
| 15 | Newton III            |          | AR1       | AR2       | AF1        | Ob             |
| 16 | Newton III            |          | AR1       | AR2       | AF1        | Ob             |
| 17 | $\Sigma F_{net} = 0$  | CI1      |           |           | CI1, G1    | AF1            |
| 18 | circ. motion          | AF1, Ob  |           | I5        | I5         | CF             |
| 19 | vel. vs acc.          | K2       | K1        | K1        | K1         |                |
| 20 | vel. vs acc.          |          | K2        | K2        |            |                |
| 21 | Newton II, s vs t     | I2       | CI3       | CI2       | I4         |                |
| 22 | Newton II             | AF4      |           | AF7       | AF6        | AF7            |
| 23 | s vs t, $\Sigma F_{net} = 0$ | I2 |           | CI3       | I2, I3     | I4             |
| 24 | $\Sigma F_{net} = 0$  |          |           | I3        |            | I3             |
| 25 | $\Sigma F_{net} = 0$  | R2       | R2        |           | R2         | CI1            |
| 26 | Newton II             | AF4      | R2, R3    | I4        | AF6        |                |
| 27 | Newton II             | AF2, R1  | I3, R1    |           | I1         | I4             |
| 28 | Newton III            |          | AF1       |           | AR1, AR2   |                |
| 28 | $\Sigma F_{net} = 0$  | Ob       |           | G1, G2    | G1         | AF3            |
| 30 | Newton II             | AF1      | I1        |           | I1         | I1             |

**Table B1**

A comparison of the dominant incorrect answers choices for the full female and male cohort (2020 to 2022 inclusive) on the pre-test, see Figure 3. In most cases the incorrect choices are identical, except for questions $Q1$, $Q5$, $Q11$, $Q13$, $Q14$, and $Q20$; as underlined.

| Gender | Q1  | Q2  | Q3  | Q4  | Q5  | Q6  | Q7  | Q8  | Q9  | Q10 |
|--------|-----|-----|-----|-----|-----|-----|-----|-----|-----|-----|
| Female | D   | B   | B   | A   | A   | A   | A   | A   | B   | D   |
| Male   | A   | B   | B   | A   | C   | A   | A   | A   | B   | D   |
|        | Q11 | Q12 | Q13 | Q14 | Q15 | Q16 | Q17 | Q18 | Q19 | Q20 |
| Female | B   | C   | B   | B   | C   | C   | A   | D   | D   | A   |
| Male   | C   | C   | C   | A   | C   | C   | A   | D   | D   | C   |
|        | Q21 | Q22 | Q23 | Q24 | Q25 | Q26 | Q27 | Q28 | Q29 | Q30 |
| Female | B   | A   | C   | C   | D   | A   | B   | D   | A   | E   |
| Male   | B   | A   | C   | C   | D   | A   | B   | D   | A   | E   |

choices $A, \ldots, E$ in the FCI), see Table B1 for the pre-test. In most cases the choices are identical except for questions $Q1, Q5, Q11, Q13, Q14$, and $Q20$. However, knowledge of student performance on all 30 questions for the largest DM (incorrect choice) contains useful conceptual information − we present them in full for completeness. From the taxonomy tables, see Tables A1 and A2, there appears to be no obvious overarching conceptual issue. However, $Q5$ and $Q11$ suggest that the female cohort was inclined towards the misconception that 'Ob − obstacles exert no force', whereas the male cohort favoured the 'I1 − impetus supplied by hit'.

*Author Biographies*

### Anna Chrysostomou

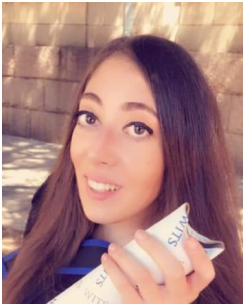

Anna is a physics student pursuing a joint PhD degree through the University of Johannesburg and the Claude Bernard University of Lyon-1. Her research lies in high-energy theoretical physics, with a specific focus on gravitational physics and its intersection with particle physics and cosmology.

### Alan S. Cornell

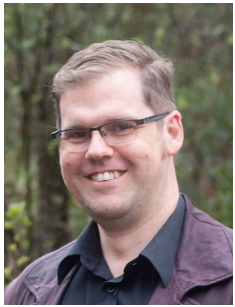

Alan is a professor of theoretical physics at the University of Johannesburg. He studied in Australia and has worked in Korea, Japan and France. His research interests are in particle and gravitational physics, and higher education pedagogy.

### Wade Naylor

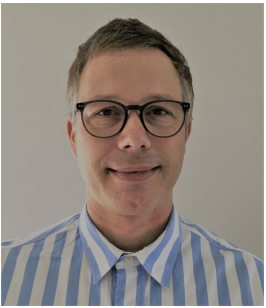

Dr Wade Naylor (MInstP) was formerly an Associate Professor of Theoretical Physics at Osaka University (2010 − 2015). After a six year career interruption to become a high school physics teacher, from 2022 Wade has been a lecturer in Physics, Mathematics & STEM education at ACU. Wade's research interests lie in Physics Education Research (PER) and research into 'misconceptions' in physics and mathematics. He is also a visiting senior researcher at the Physics Dept., University of Johannesburg (UJ).